\documentclass[prl, twocolumn, superscriptaddress, nofootinbib, nobibnotes]{revtex4-2}

\usepackage[english]{babel}
\usepackage{hyperref}

\usepackage{soul}

\usepackage[pdftex]{graphicx}
\usepackage{color}
\usepackage[toc,page]{appendix}
\usepackage{float}
\usepackage{amsmath}
\usepackage{amssymb}
\usepackage{mathtools}
\usepackage{braket}

\usepackage[sort&compress]{natbib}

\newcommand{\un}[1]{\ensuremath {\,\mathrm{#1}}}
\renewcommand{\v}[1]{\ensuremath {\boldsymbol{#1}}}
\renewcommand{\vec}[1]{\ensuremath {\boldsymbol{#1}}}

\newcommand{\fig}[1]{Figure~\ref{fig:#1}}

\newcommand{\eq}[1]{(\ref{eq:#1})}
\newcommand{\lr}[1]{\ensuremath{\left( #1 \right)}}

\renewcommand{\Re}[1]{\ensuremath{\mathrm{Re} \left(#1\right)}}
\renewcommand{\Im}[1]{\ensuremath{\mathrm{Im} \left(#1\right)}}
\newcommand{\I}{\mathrm{i}}

\newcommand{\Sg}{\Sigma}

\newcommand{\Tr}[1]{\ensuremath{\mathrm{Tr}\left(1\right)}}

\usepackage{ulem}

\begin{document}

\title{Electronic transport and anti-super-Klein tunneling in few-layer black phosphorous}

\author{Jorge Alfonso Lizarraga-Brito}
\email{jorge\_lizarraga@icf.unam.mx}
\affiliation{Instituto de Ciencias F\'isicas, Universidad Nacional
  Aut\'onoma de M\'exico, Cuernavaca, Mexico}

\author{Armando Arciniega-Gutiérrez}
\email{armandoarciniega@ciencias.unam.mx}
\affiliation{Instituto de Ciencias F\'isicas, Universidad Nacional
  Aut\'onoma de M\'exico, Cuernavaca, Mexico}

\author{Yonatan Betancur-Ocampo}
\email{ybetancur@fisica.unam.mx}
\affiliation{Instituto de F\'isica, Universidad Nacional
  Aut\'onoma de M\'exico, Ciudad de México, Mexico}

\author{Thomas Stegmann}
\email{stegmann@icf.unam.mx}
\affiliation{Instituto de Ciencias F\'isicas, Universidad Nacional
  Aut\'onoma de M\'exico, Cuernavaca, Mexico}

\date{\today}

\begin{abstract}
The electronic transport in few-layer black phosphorus (FLBP) nanoribbons is studied theoretically. The system is modeled on the basis of band-structures, which have been measured recently by $\mu$-ARPES experiments. We show that the anisotropic bands of FLBP leads to highly anisotropic transport properties; while the current in one direction can be rather focused, it can be strongly disperse in the orthogonal direction. The low-energy current is carried mainly in the central layer due to the vertical confinement of the electrons. In FLBP pn junctions, generated by the electrostatic potential of a gate contact in a certain region of the system, the electrons pass through the interface of the junction, if it is oriented along the zigzag direction of FLBP. If the junction is rotated by 90 degree and oriented along the armchair direction, the current is reflected completely for all angles of incidence and for a wide range of electron energies. This omni-directional total reflection is named anti-super-Klein tunneling as it is due to opposite pseudo-spins of the electrons in the two region of the pn junction. The effect of oxidation of the top layer of FLBP pn junctions is investigated and it is found that, while the current flow in the top layers is strongly suppressed, the anti-super-Klein tunneling persists.
\end{abstract}

\maketitle

\section{Introduction}

Phosphorene, the mono-layer of black phosphorous, has received in recent years many attention due to its interesting properties like an intrinsic band-gap and high electron mobility \cite{kou2015phosphorene, Carvalho2016, Zhang2024, Ding2024}. In addition, its anisotropic band structure, where the electrons behave like massive Dirac fermions in a certain crystallographic direction, whereas they behave as a Schr\"odinger electrons in the orthogonal direction, as well as effects like negative reflection and the anti-super-Klein tunneling make phosphorene a promising 2D material for nanoelectronic applications \cite{de2017anisotropic, li2017direct, ehlen2016evolution, peng2014strain, taghizadeh2015scaling, taghizadeh2016strain, ezawa2014topological, midtvedt2017multi, voon2015effective, brown1965refinement, rudenko2014quasiparticle, rudenko2015toward, ccakir2014tuning, elahi2015modulation, li2018tuning, Betancur2019, Betancur2020}.

An obstacle in the usage of the material is the fact that often not a single layer but systems composed of several layers -- named few layer black phosphorous (FLBP) --  are synthesized. FLBP itself is of interest, for example, because its electronic properties can be rather identical to those of the monolayer and can be tuned by the number of layers, the way they are stacked, doping, external pressure and a twist between the layers \cite{dai2014bilayer, kim2015observation, gong2016hydrostatic, kim2017microscopic, xu2018role, fujii2020pressure, Margot2023, gao2024unified, soltero2022moire}. This makes FLBP a promising material for device applications. For example, it has been used to create high-performance field-effect transistors (FETs) \cite{Li2014, Liu2014, Doganov2015, feng2018high, feng2018complementary, ameen2016few, liu2015device}, which can even be thickness-engineered to outperform some previous designs \cite{chen2016thickness}. Additionally, its ability to operate effectively in the gigahertz range makes it a strong candidate for radio-frequency transistors \cite{wang2014black}, while the high carrier mobility enhances its potential for use in photovoltaic energy conversion devices \cite{deng2014black}.

However, despite these exciting properties and potential applications, an obstacle remains, the oxidation of the phosphorene layers, which can occur within hours under ambient conditions. This rapid degradation limits the material's stability, making it essential to develop effective methods for passivation or encapsulation to protect the phosphorene surface from oxygen exposure \cite{wood2014effective, pei2016producing, tayari2015two, island2015environmental, ziletti2015oxygen, doganov2015transport, island2015environmental, sang2019recent}.

Not all aspects are negative when it comes to the high reactivity of phosphorene. It can even be seen as beneficial, as it allows for the tailoring of its properties. For instance, the oxidation of phosphorene can be used to tune the band gap by reducing the effective thickness of the system \cite{ziletti2015phosphorene, kwon2016ultrathin}. Also, when using it as a battery material, oxidized black-phosphorus-based electrodes present a capacitance exceeding significantly that one of pristine black phosphorous \cite{nakhanivej2019revealing}, or using actively oxidized black phosphorous nanosheets to obtain a (nearly) zero friction surface, known as super-lubricity \cite{ren2021superlubricity}.

\begin{figure*}[t]
  \centering
  \includegraphics[scale=0.6]{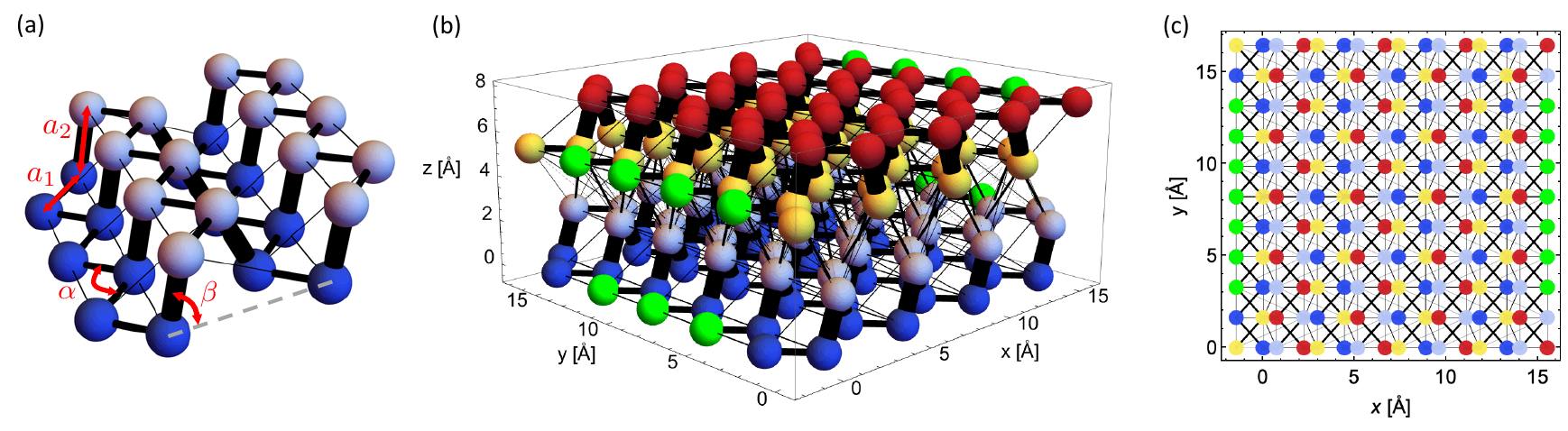}
  \caption{(a) Phosphorene nanoribbon in order to define the lattice constants $a_{1/2}$ and angles $\alpha, \beta$. (a) Few-layer black phosphorus nanoribbon of $N=2$ stacked layers of phosphorene. The color of the phosphorous atoms indicates their height in the $z$ direction. The atoms at the edges where the contacts are attached are marked in green. (b) Projection of the atoms in the $xy$ plane showing that each phosphorene layer can be interpreted as a strongly deformed hexagonal lattice with an armchair edge in the $x$ direction and a zigzag edge in the $y$ direction.}
  \label{fig:1}
\end{figure*}

In this paper, we investigate theoretically the electronic transport in FLBP. We use a tight-binding Hamiltonian with parameters from recent $\mu$-ARPES experiments \cite{Margot2023, Margot2025} and employ the Green's function method to calculate the current flow. We show that the anisotropic bands of FLBP leads to highly anisotropic transport properties; while the current in one direction can be rather focused, it can be strongly disperse in the orthogonal direction. The low-energy current is carried mainly in the central layer. In particular, we investigate the current flow in FLBP pn junctions, generated by the electrostatic potential of a gate contact in a certain region of the system. We show that the anti-super-Klein tunneling, reported previously in phosphorene \cite{Betancur-Ocampo2019, Betancur2020}, persists in FLBP. This phenomena means total reflection of the electrons for all incidence angles at the surface of pn junction oriented along the zigzag direction of the phosphorene lattice. It is a consequence of the anisotropy of the band structure of phosphorene and the pseudo-spin structure of the wave function of the electrons. We also investigate the effect of the oxidation of FLBP, removing randomly up to 30\% of the bonds of the outermost layer and find that the current flow in the outer layers is suppressed strongly while it remains largely unaffected in the inner layers. Most importantly, the anti-super-Klein tunneling is not altered by the oxidation of the material.

\section{Device model \& computational methods}

The system, which is sketched in \fig{1}, consists of $N$ layers of phosphorene stacked upon each other with AB stacking symmetry which is energetically favorable and found in the black phosphorous crystal. The individual layers have a distance of $0.53 \un{nm}$ \cite{Chaudhary2022}. As phosphorene itself is a buckled material, we have to deal effectively with $2N$ layers. The phosphorene layers have lattice constants $a_1= 2.22\un{\AA}$, $a_2=2.24 \un{\AA}$ and the angles $\alpha= 96.5^\circ$, $\beta= 72^\circ$. Edges in the $x$ and $y$ direction have the armchair and zigzag shape, respectively. The slight changes of the lattice constants with the number of phosphorene layers are not considered here. 

The electronic structure of the system is described by a tight-binding model  \begin{equation}
  \label{eq:1}
  H= \sum_{n=1}^N H_n + \sum_{n=1}^{N-1} H_n^{\perp},
\end{equation}
where
\begin{equation}
    H_n= \sum_{(i,j) \leq {4}} t_{(i,j)} \ket{i^n}\bra{j^n} + \text{H.c.}
\end{equation}
describes the interaction within the $n$th phosphorene layer, while 
\begin{equation}
H_n^\perp= \sum_{(i,j) \leq 4} t_{(i,j)
}^\perp \ket{i^n}\bra{j^{n+1}} + \text{H.c.}
\end{equation}
gives the interaction between the layers. Intra and intelayer couplings are taken into account up to fourth nearest neighbors. The parameters have been obtained on the basis of recent $\mu$-ARPES experiments \cite{Margot2023, Margot2025}%
\footnote{The parameters, which were provided kindly by Florian Margot, differ from those published originally in Refs. \cite{Margot2023} due to a minor error in the tight-binding model (equation (4d) in the Supplementary Material of Refs. \cite{Margot2023} should read $ r_4^{||}= (-(2a_1 \cos(\alpha/2)+ a_2 \cos(\beta)), 0)$ and equation (7a) $r_1^\perp = (a_2 \cos(\beta), \pm a_1 \sin(\alpha/2)) $). Corrected parameters have been published while finishing this manuscript \cite{Margot2025}.}
and are summarized in Table \ref{tab:1}.

\begin{table}[htb]
\begin{tabular}{||cc||cc||l}
\cline{1-4}
\multicolumn{2}{||c||}{$||$ [eV]} & \multicolumn{2}{c||}{$\perp$ [eV]} &  \\ \cline{1-4}
\multicolumn{1}{||l|}{$t_1$} & -1.494 & \multicolumn{1}{l|}{$t_1^\perp$} & 0.510   &  \\ \cline{1-4}
\multicolumn{1}{||l|}{$t_2$} & 3.630  & \multicolumn{1}{l|}{$t_2^\perp$} & 0.270   &  \\ \cline{1-4}
\multicolumn{1}{||l|}{$t_3$} & -0.260 & \multicolumn{1}{l|}{$t_3^\perp$} & -0.087 &  \\ \cline{1-4}
\multicolumn{1}{||l|}{$t_4$} & 0.364  & \multicolumn{1}{l|}{$t_4^\perp$} & -0.093  &  \\ \cline{1-4}
\end{tabular}
\caption{Tight-binding parameters up to fourth nearest neighbors within the phoshorene layers ($||$) and between the layers ($\perp$). Data are on the basis of Refs. \cite{Margot2023, Margot2025}$^1$.}
  \label{tab:1}
\end{table}

\begin{figure}
    \centering
    \includegraphics[width=0.98\linewidth]{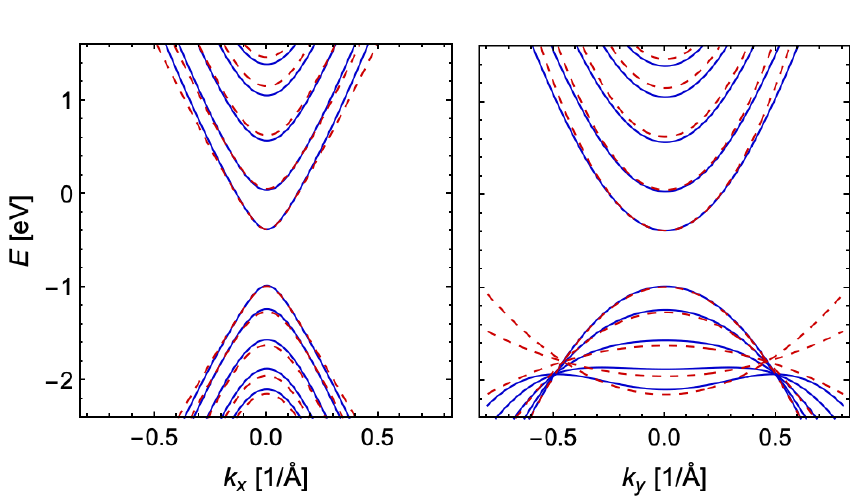}
    \caption{Band structure of 5-layer FLBP. Solid-blue lines: numerical calculated bands using the tight-binding Hamiltonian \eq{1}. Dashed-red lines: energy bands from the simplified low-energy Hamiltonian \eq{4}. The two panels show the band structure in the $k_x$ and $k_y$ direction, demonstrating its strong anisotropy. The simplified model describes well the low energy bands close to the band gap, which is the relevant energy region for this study.}
    \label{fig:2}
\end{figure}

A plane-wave ansatz (see the Appendix II for details) leads to the energy bands that are depicted in \fig{2} (solid-blue curves) for a system consisting of $N=5$ layers. The energy bands are highly anisotropic with respect to the $k_x$ and $k_y$ directions. Moreover, the conduction and valence bands in the $k_y$ direction are highly asymmetric; the valence bands are much flatter than the conduction bands. The tight-binding model describes correctly the reduction of the band gap with the number of layers, which is observed in several experiments \cite{Margot2023}.

In order to obtain a continuous Hamiltonian for the low-energy bands of FLBP, we follow Ref. \cite{Sousa2017} and apply Toeplitz theory to block diagonalize the Hamiltonian, decouple \textit{approximately} the high from the low-energy bands and expand the Hamiltonian around the $\Gamma$ point (see the Appendix II for details). Finally, we arrive at the following effective Hamiltonian 
\begin{widetext}
    \begin{equation}
    \label{eq:4}
        H_{n}^\text{red}\approx \left(\delta_{n}+\frac{k^2_x}{2\mu^{n}_x}+\frac{k^2_y}{2\mu^{n}_y}\right)\sigma_0 +\left(\Delta_{n}+\frac{k^2_x}{2m^{n}_x}+\frac{k^2_y}{2m^{n}_y}\right)\sigma_x + v^{n}_xk_x\sigma_y,
    \end{equation}
\end{widetext}
for the $n=1 \dots N$ layers. The effective mass terms read
\begin{align}
    \frac{1}{m^{n}_x} & \equiv -2a^2_1t_1\cos^2(\alpha/2)-a^2_2t_2\cos^2(\beta) - t_4l^2_2\nonumber\\
&\quad -2\lambda_n[t^\perp_1a^2_1\cos^2(\beta)+t^\perp_4l^2_2],\\
\frac{1}{m^{n}_y} &\equiv -2a^2_1[t_1+\lambda_n(t^\perp_1  + t^\perp_4)]\sin^2(\alpha/2),\\
\frac{1}{\mu^{n}_x} & \equiv -2\lambda_n l^2_1 (t^\perp_2+2t^\perp_3),\\
\frac{1}{\mu^{n}_y} & \equiv -8(t_3+2t^\perp_3\lambda_n)a^2_1\sin^2(\alpha/2),
\end{align}
while the velocity and semi-gaps are given by
\begin{align}
v^{n}_x &\equiv -2a_1t_1\cos(\alpha/2) +a_2t_2\cos(\beta)-t_4l_2 \notag\\
&\quad+2t^\perp_1 a_1\lambda_n \cos(\beta)-2t^\perp_4\lambda_nl_2,\\
\Delta_{n} &\equiv 2t_1+t_2+t_4 + 2\lambda_n(t^\perp_1+t^\perp_4),\\
\delta_{n} &\equiv 2t_3+\lambda_n(2t^\perp_2+4t^\perp_3),
\end{align}
with 
\begin{equation}
\label{eq:lambda}
    \lambda_n \equiv \cos\left(\frac{2n-1}{2M+1}\pi\right)
\end{equation}
and $l_1 \equiv a_1\cos(\alpha/2)+a_2\cos(\beta)$, $l_2 \equiv 2a_1\cos(\alpha/2)+a_2\cos(\beta)$. Note that \eq{lambda}, which we found empirically, differs from the one given in Refs. \cite{Sousa2017} but leads to slightly better agreement to the numerically calculated energy bands (see the Appendix II). The energy bands from this low-energy model are shown in \fig{2} (red-dashed curves). In the $k_x$ direction the gapped Dirac cones of ultra-relativistic particles are found, while in the $k_y$ direction the gapped parabolic bands of non-relativistic electrons are obtained. In general, good agreement between the effective model and the numerical calculated bands is found, in particular for the low energy bands close to the gap that will be most relevant for this study. We will use the effective model later to understand some transport features of few-layer black phosphorous. 


The electron transport in FLBP is studied by means of the non-equilibrium Green's function (NEGF) method. As detailed introductions can be found in various text books \cite{Datta2005, Datta1997, DiVentra2008}, we summarize here only briefly the essential equations. The Green's function of the system is given by 
\begin{equation}
  \label{eq:9}
  G(E)= \lr{E-H-\Sg_S -\Sg_D}^{-1},
\end{equation}
where $E$ is the energy of the injected electrons and $H$ is the tight-binding Hamiltonian, \eq{1}. The self-energies $\Sg_{S/D}$ describe the effect of the source and drain contacts and are modeled within the wide-band model 
\begin{equation}
  \label{eq:12b}
  \Sg_{S/D}=  \sum_{i \in \text{contact}} -\I \, \eta\ket{i} \bra{i},
\end{equation}
representing contacts with a constant, energy independent surface density of states, $\text{DOS} \propto \eta = \text{const}$. The sum is over those phosphorous atoms that are connected to the contacts, see the green marked atoms in \fig{1}~(b). In the following we set $\eta=t_1^\text{ml}$, though our results do not depend crucially on that choice. The transmission between the contacts is given by
\begin{equation}
    T_{SD}(E)= \text{Tr}(\Gamma_S G \Gamma_D G^\dagger)
\end{equation}
and represents the conductance $\mathcal{G}= (2e^2/h)T_{SD}$ for electrons with energy $E$. The inscattering functions are defined as $\Gamma_{S/D}= -2 \text{Im}(\Sigma_{S/D})$. Finally, the current flowing between the atoms at positions $\v r_i$ and $ \v r_j$ is calculated by
\begin{equation}
  \label{eq:13}
  I_{ij} = \frac{2e}{h} \textrm{Im}(t_{ij}^*\, (G\, \Gamma_S\, G^{\dagger})_{ij}).
\end{equation}

\section{Results}


\begin{figure}[t]
  \centering
  \includegraphics[scale=0.266]{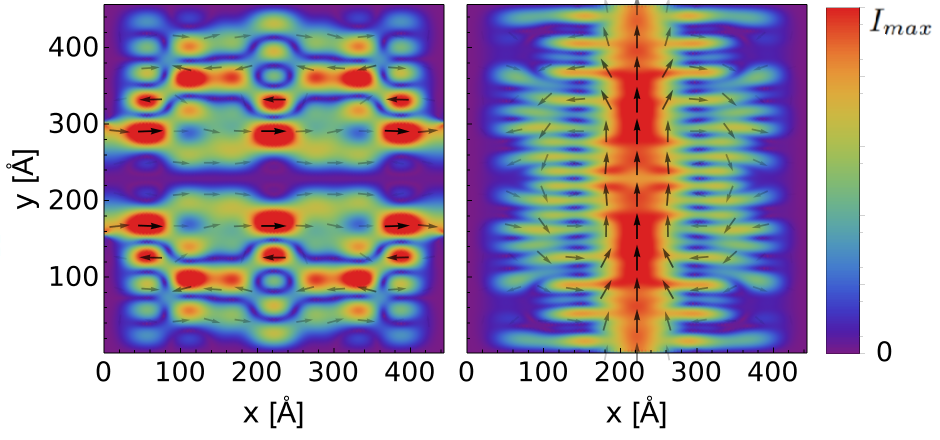}
  \caption{Local current flow in the phosphorene monolayer for electrons with energy $\mu = 150 \un{meV}$ above the conduction band edge. The
  current vector field is given by the arrows, its density by the color shading. The current is injected at the left edge (a) and the bottom edge (b).}
  \label{fig:3}
\end{figure}

We start by analyzing the local current flow in phosphorene, see \fig{3}. The current is injected at $\mu=-0.1 t_1^{||} = 150 \un{meV}$ above the conduction band edge. The arrows represent the current vector field and the color shading the current density. Like the energy bands, the current flow is highly anisotropic; while a rather narrow electron beam can be observed when the current flows in the $y$ direction (established by a pair of contacts at the bottom and top edges), the current is much more disperse in the $x$ direction (generated by a pair of contacts at the left and right edges). Moreover, a ripple pattern is found in the current density which can be explained by the finite size of the system ($450 \times 450 \un{\AA} $) and the reflections taking place at the edges of the system (even in presence of the contacts), inducing the standing waves of a quantum well.


\begin{figure*}[t]
  \centering
  \includegraphics[scale=0.26]{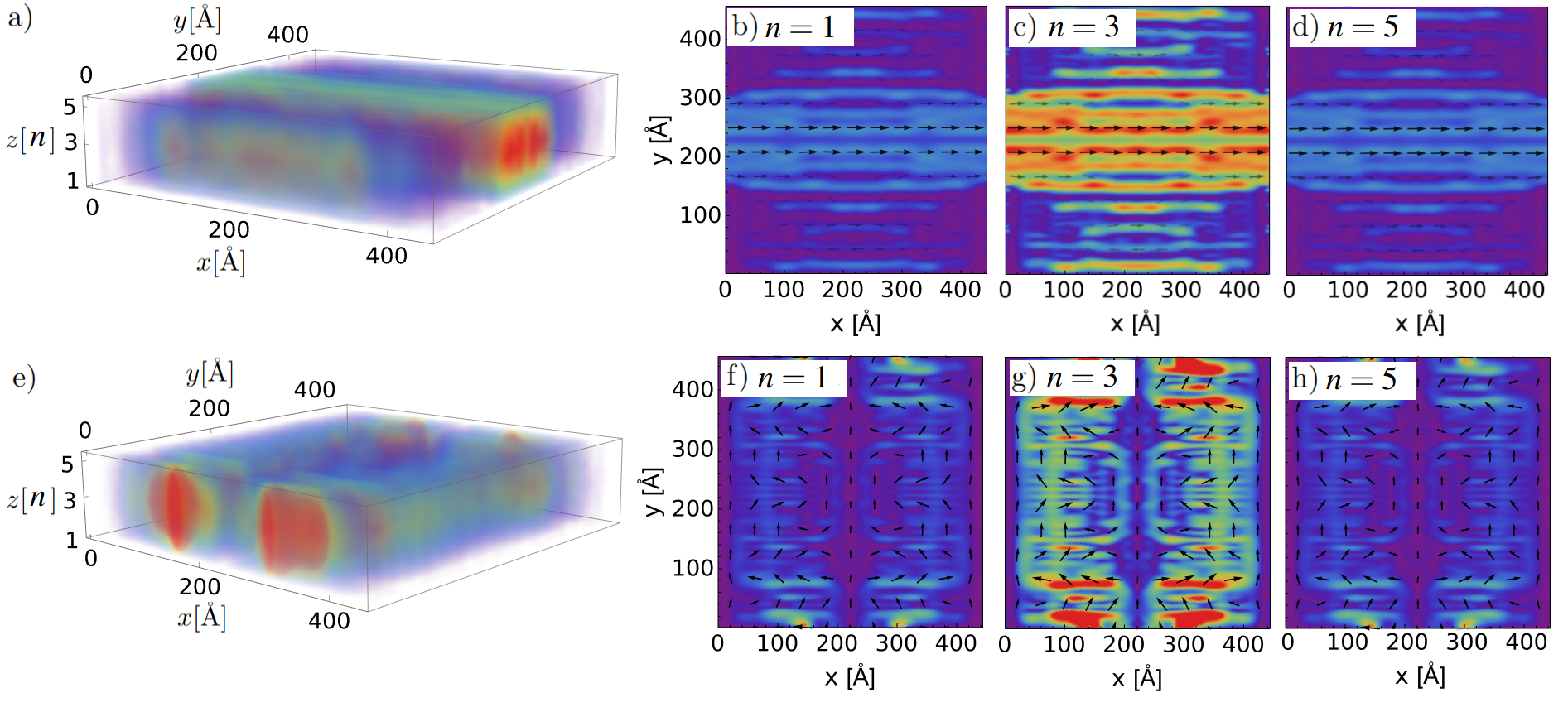}
  \caption{Local current flow in a few-layer black phosphorus nanoribbon consisting of 5 layers. Electrons are injected at $\mu = 150 \un{meV}$ above the conduction band edge. The contacts are either along the $y$ direction (a-d) or along the $x$ direction (e-h). Panels (a, e) show 3D figures of the current density, panels (b-d, f-h) the current in certain layers. Due to the anisotropy of the energy-bands of FLBP, the current flow patterns also depend strongly on the crystallographic direction. The current is concentrated in the central layer, which is to be expected for confined electrons at low energies.}
  \label{fig:4}
\end{figure*}

When we go from the monolayer to a FLBP system consisting of $N=5$ layers, we find the current flow patterns shown in \fig{4}. As in the monolayer, the current is injected again at the zigzag (a) and armchair edge (e) at $\mu = 150 \un{meV}$ above the conduction band edge. In the $x$ direction, it can be seen that electrons tend to move mainly in a concentrated beam from left to right, while in the $y$ direction the current is split into two beams. However, note that the distribution of the current depends also strongly on the system size and electron energy. The important point that we want to make here is that the local current flow in FLBP is highly anisotropic. In all cases, the electron current is mainly concentrated in the central layer ($n=3$). This vertical confinement can be understood also by the properties of a quantum well, whose low energy modes favor the localization in the central layer.


\begin{figure}[t]
  \centering
  \includegraphics[scale=0.45]{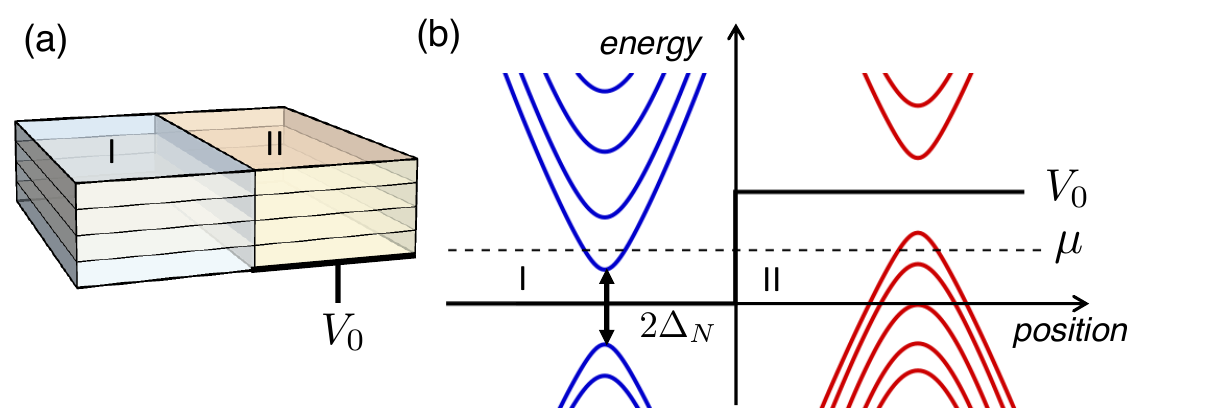}
  \caption{A transverse electric field is generated in region II of the FLBP system through a gate contact. The associated electrostatic potential $V_0$ shifts the energy bands in region II with respect to those in region I. In this way, the electron with energy $\mu$ can go from the conduction to the valence band (dashed horizontal line).}
  \label{fig:5}
\end{figure}

Having understood qualitatively the local current flow in FLBP nanoribbons, we investigate in the next step how it can be controlled by means of a transverse electric field, generated by a gate contact. We consider the setup shown in \fig{5}, where the gate contact is present only in a certain region of the system (region II). The electric field shifts through its electrostatic potential $V_0$ the energy bands in this region and can generate a FLBP pn junction, where the electrons change from the conduction band to the valence band, when going from region I to region II. The electrostatic potential is modeled by
\begin{equation}
 V_n= V_0\lr{1+\frac{\nu}{2}(2n-1-N)},
\end{equation}
where the parameter $\nu$ takes into account a linear change of the potential within the different phosphorene layers. Additionally, we consider that the potential increases linearly when going from region I to region II, over the range of 2 unit cells. 

\begin{figure*}[t]
  \centering
  \includegraphics[scale=0.31]{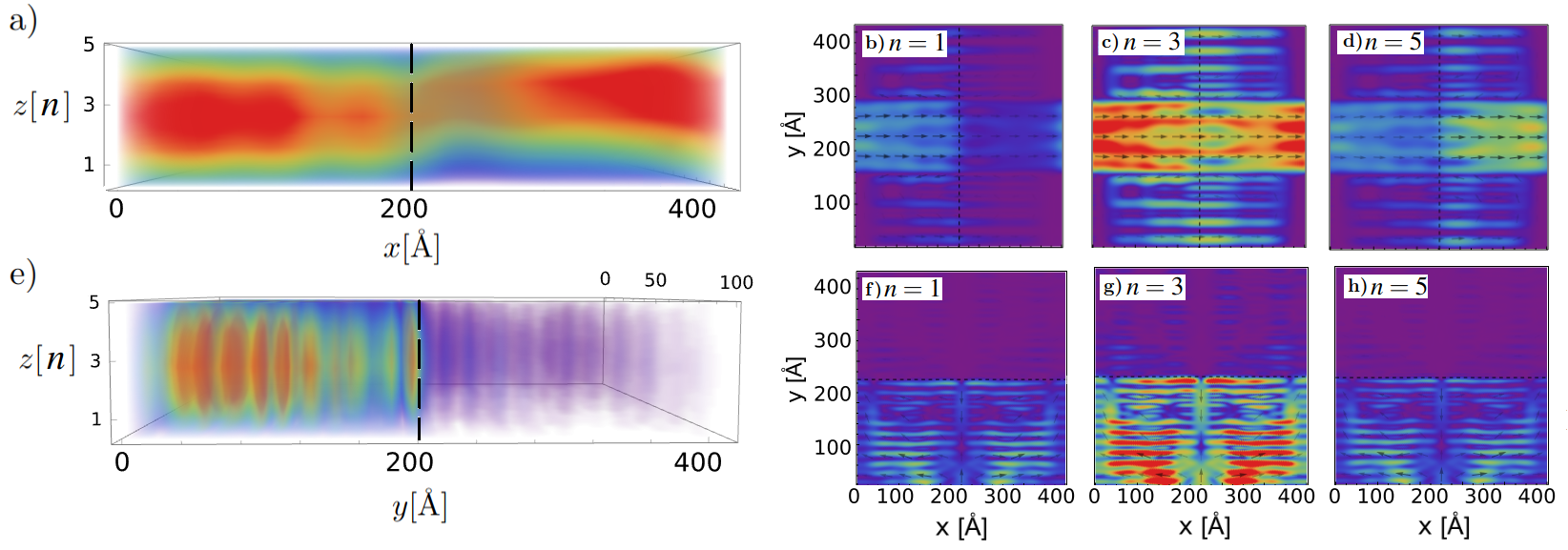}
  \caption{Current flow in a few-layer black phosphorus pn junction with $N=5$ layers and $V_0=2\mu$. If the interface of the junction is aligned parallel to the zigzag edge (a-d), the current is passing through the junction. If the junction is parallel to the armchair edge (e-h) the current is almost completely reflected due to anti-super-Klein tunneling. In (e-h) the electrons are injected at $\mu=150\un{meV}$ (above the conduction band edge), while in (a-d) $\mu=50 \un{meV}$ is used to show the deviation of the current towards the top layer due to the transverse electric field ($\nu=0.1$).}
   \label{fig:6}
\end{figure*}

The current density of such a FLBP pn junction with $V_0=2\mu$ is shown in \fig{6}. We observe that the current flow depends strongly on the orientation of the underlying phosphorene lattice. When the electrons are injected at the zigzag edge (panels a-d) the current is passing through the junction. In contrast, when the current is injected at the armchair edge (panels e-h) the current is (almost) completely reflected. This phenomena -- omni-directional total reflection of the current at the interface of the pn junction aligned parallel to the armchair edge -- has been named anti-super-Klein tunneling, and has been reported previously for phosphorene \cite{Betancur2019}. Here, we demonstrate that the effect is observed also in FLBP pn junctions. Note that, we have considered that the electrostatic potential changes 10\% between the layers ($\nu=0.1$), which we estimate as an almost unrealistic high value in order to test the validity of the anti-super-Klein tunneling. Like in the case without electrostatic potential, the main current is carried in the central layer. In the case of the zigzag injection (panel a-d) we have used a rather low electron energy, $\mu=80\un{meV}$ to show that the current is deviated towards the top layer due to the transverse electrical field.

Like in the monolayer, the anti-super-Klein tunneling is not due to an energetically forbidden region (a band gap) but can be understood in terms of the pseudo-spin structure of the electron wavefunction in FLBP. The direction of the pseudospin in each layer is given by
\begin{equation}
    \phi^{(n)}(\vec{k}) = \arctan\left(\frac{v^{n}_x k_x}{\Delta_{n}+\frac{k^2_x}{2m^{n}_x}+\frac{k^2_y}{2m^{n}_y}}\right),
\end{equation}
which is anisotropic due to the dispersion relation of the Hamiltonian \eq{4}.

\begin{figure}
    \centering
    \begin{tabular}{c}
         \includegraphics[width=0.97\linewidth]{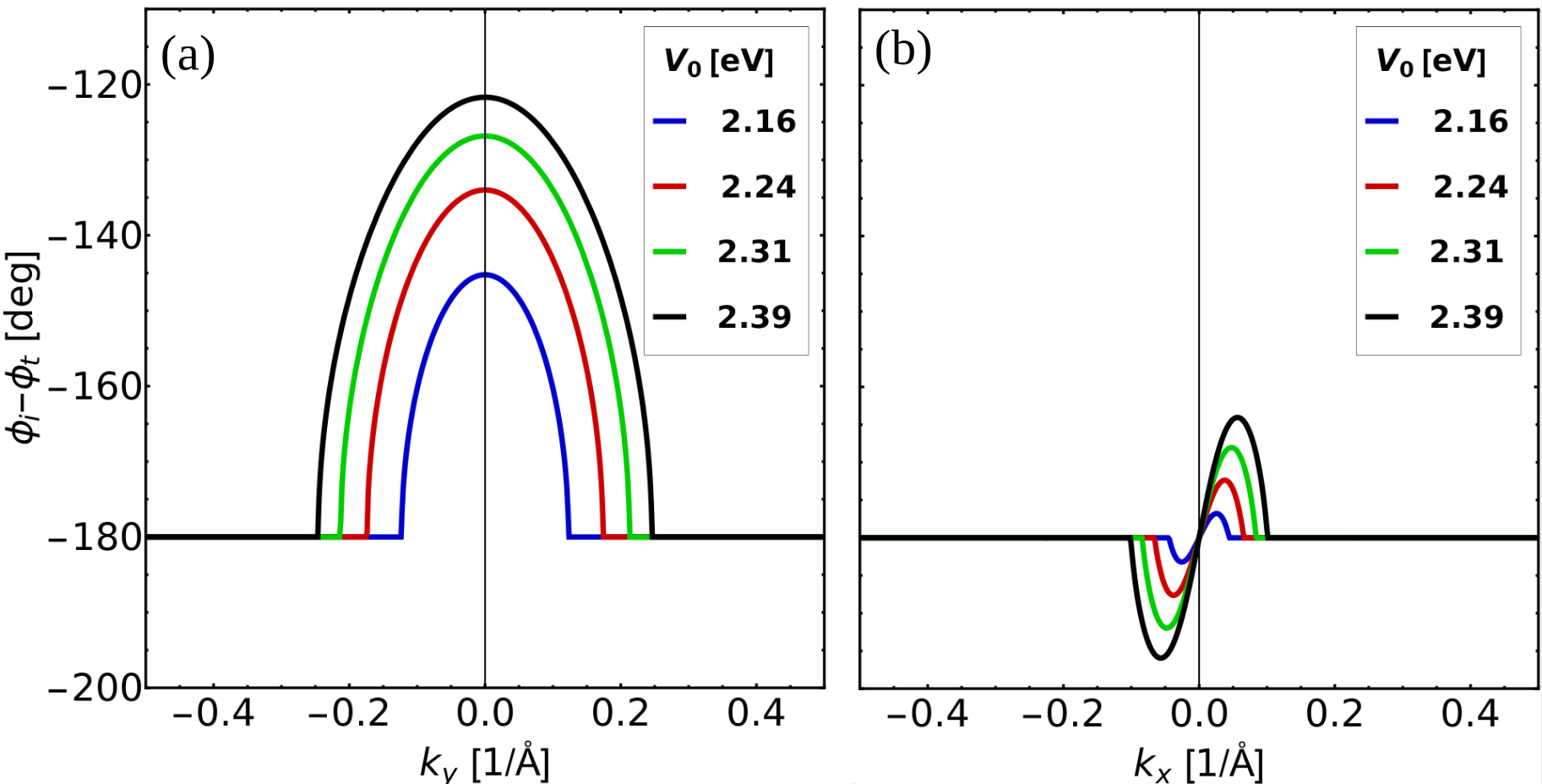}
    \end{tabular}
    \caption{Difference of the pseudo-spin between regions I and II as a function of the conserved wave vector component $k_x$ and $k_y$ for a few-layer black phosphorus nanoribbon with $N = 5$ layers, which corresponds to the case of a current in $x$ direction (a), and $y$ direction (b). The electron energy is $\mu = 150 \un{meV}$ and colored curves are shown for different values of $V_0$.}
    \label{fig:7}
\end{figure}

The difference of the pseudo-spin between region I and II is shown in \fig{7} as a function the wave vector component that is parallel to the interface of the pn junction and thus, conserved. This figure proves that the pseudo-spins are almost antiparallel in the $y$ direction (panel b, $k_x $ conserved) for a wide range of different electrostatic potentials, which move the electrons from the valence to the conduction band. In the $x$ direction (panel a, $k_y$ conserved) their difference is much less pronounced. As the pseudo-spin is a conserved quantity at the interface of the pn junction, this explains the anti-super-Klein tunneling in the $y$ direction, while transport is possible in the $x$ direction.

\begin{figure*}[t]
  \centering
  \includegraphics[scale=0.5]{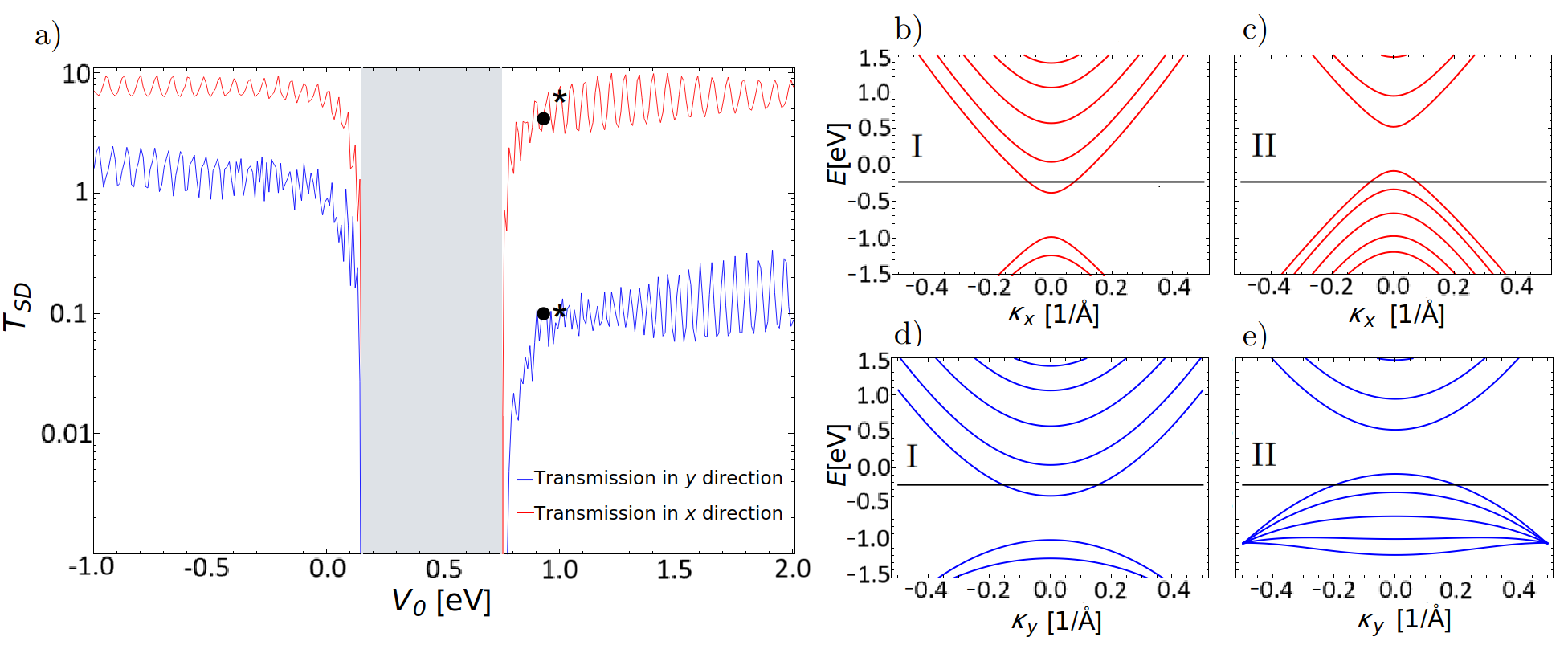}
  \caption{(a) Transmission $T_{SD}$ (in logarithmic scale) in a few-layer balck phosphorus pn junction (5 layers) as a function of the electrostatic potential $V_0$. Electrons injected at energy $\mu= 150 \un{meV}$ above the conduction band edge. The gray shaded region shows the intrinsic band gap of FLBP where transport is naturally blocked. The transmission to the left and right of this region is rather symmetric for transport in the $x$ direction (red curve). In the $y$ direction (blue curve) the transmission between left and right differs by an order of magnitude due to anti-super-Klein tunneling. The star symbol marks the potential where more than two energy bands are occupied in region II, while the dot symbol highlights the symmetric case $V_0= 2\mu$. (b-e) Energy bands in region I and II in the two orthogonal directions of transport.}
  \label{fig:8}
\end{figure*}

In order to study further the robustness of the anti-super-Klein tunneling, we show in \fig{8} (a) the transmission $T_{SD}$ as a function of the potential $V_0$ for electrons injected at energy $\mu= 150 \un{meV}$ above the conduction band edge in a system of 5 layers. The transmission in the $x$ direction (junction parallel to the zigzag edge) is shown by the red curve while the transmission in the $y$ direction (junction parallel to the armchair edge) is represented by the blue curve. The corresponding energy bands in the two directions and two regions are shown in panels (b-e). The energy of the injected electrons is highlighted by a horizontal line. For values $V_0>754\un{meV}$ the electrons go from the conduction to the valence band when crossing the interface of the pn junction (see panels (b-e)), while for $V_0<151\un{meV}$ they remain in the conduction band. For values $151<V_0<754 \un{meV}$, see the gray shaded region, the transport is blocked in both directions due to the intrinsic band gap of FLBP. In Appendix I, the interested reader can find the transmission in the case of a phosphorene pn junction, demonstrating also in this case the anti-super-Klein tunneling for a broad parameter range.

In the $x$ direction the transmission to the left and right of the band gap are rather similar. In contrast, in the $y$ direction the transmission is strongly asymmetric; to the left of the gray-shaded region it is much larger than to the left, where it decreases by one order of magnitude (note the logarithmic scale of the transmission). This asymmetry is due to the anti-super-Klein tunneling, which therefore is not only observed under the condition $V_0=2\mu$ (this condition is marked by a black dot in panel a and shown in panels b-e) but under the much general condition that the electron switches between the conduction and valence bands. The persistence of the anti-super-Klein tunneling can be explained by the fact that the direction of the pseudo-spin varies only slowly with the potential $V_0$ and remain approximately anti-parallel for different values of $V_0$, see \fig{7}. The anti-super-Klein tunneling is neither limited to low energies, where only a single band is occupied (as shown in panels b-e) but persists also when more energies bands are occupied, see the small star in \fig{8} (a) indicating the potential from which on more than two energy bands are occupied (in region II).


\begin{figure*}[t]
  \centering
  \includegraphics[scale=0.29]{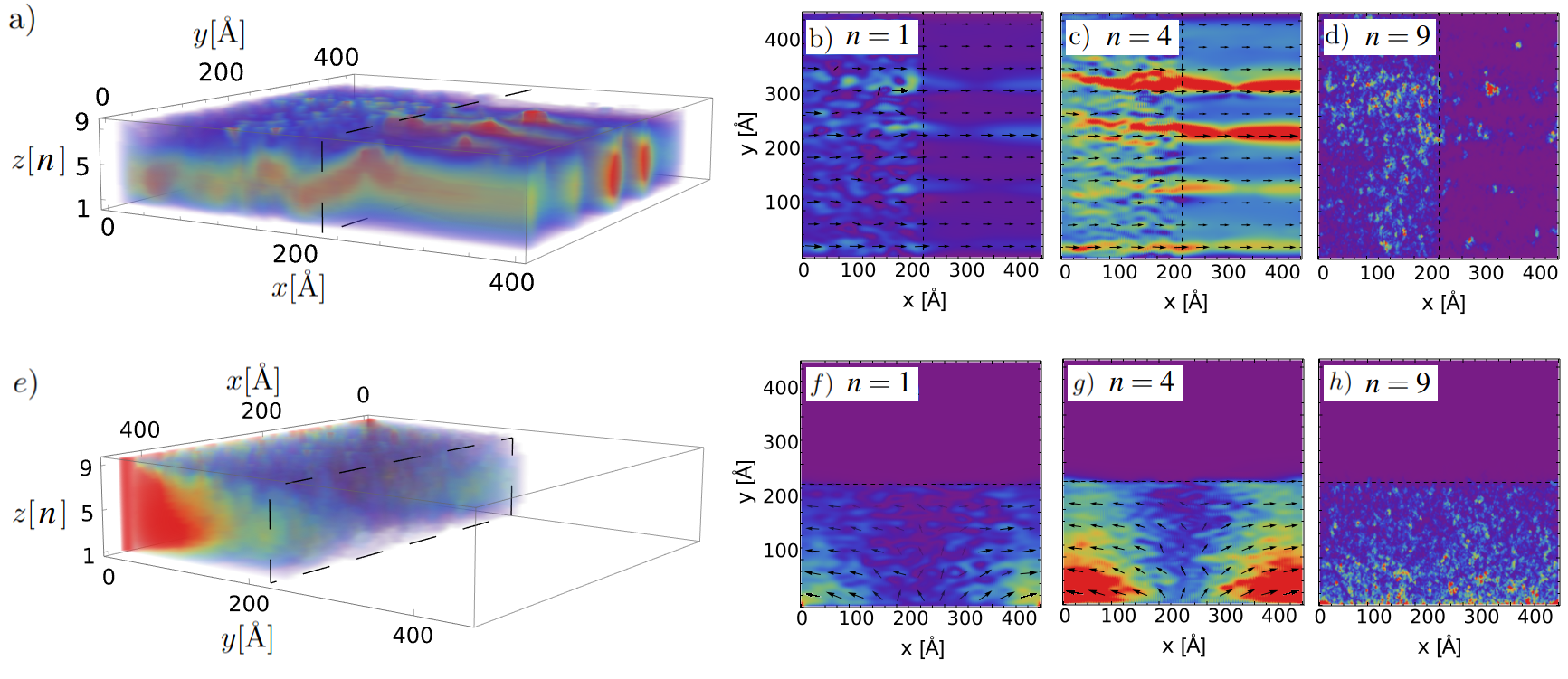}
    \caption{Local current flow in a FLBP pn junction (9 layers, $\mu=100 \un{meV}$, $V_0 =1.6\mu$). In order to model the effects of oxidation, we have deleted randomly 30 \% of the bonds of the top layer and performed an ensemble average over 20 realizations. The current flow in the top layer is strongly reduced by the disorder, while it persist in the central and bottom layers. The anti-super-Klein tunneling (e-h) are observed clearly.}  
  \label{fig:9}
\end{figure*}

As phosphorene tends to oxide, we finally investigate how the electron transport is affected by defects at the surface of the system, modeled by deleting randomly a certain degree of the bonds of the top layer. In \fig{9}, we show the local current in a FLBP pn junction of 9 layers. 30 \% of the bonds of the top layer have been deleted and an average over 20 realizations has been performed. We observe that the current in the top layer is strongly reduced by the defects, while it persists in the central and bottom layers. Most importantly, the anti-super-Klein tunneling can be observed clearly. Note that in contrast to \fig{3}, the contacts in \fig{9} extend over the full edge of the nanoribbon.

\begin{figure}[t]
  \centering
  \includegraphics[scale=0.43]{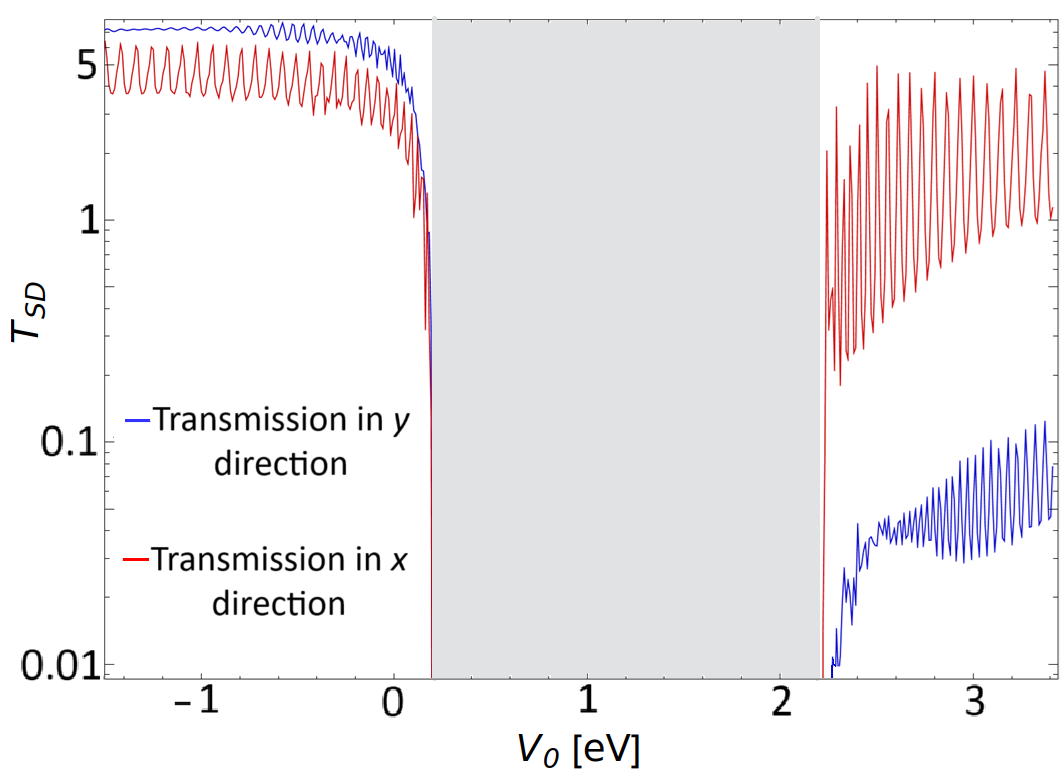}
  \caption{Transmission $T_{SD}$ (in logarithmic scale) in a phosphorene pn junction as a function of the electrostatic potential $V_0$. Electrons are injected at $\mu= 200 \un{meV}$ above the conduction band edge. The figure proves (like \fig{8} for FLBP) that anti-super-Klein tunneling is found in the $y$ direction, if the electrostatic potential of the pn junction moves to electrons form the conduction to the valence band (or vice versa).}  
  \label{fig:10}
\end{figure}

\section{Conclusions}

In this paper, we have investigated the electronic transport in FLBP nanoribbons. The system has been modeled by a tight-binding Hamiltonian that takes into account up to fourth nearest neighbors within and in between the phosphorene layers. The tight-binding parameters have been obtained by means of $\mu$-ARPES measurements of the band structure of FLBP \cite{Margot2023}. The parameters reported in Table~\ref{tab:1} differ from those in the original publication due to an error in their tight-binding model which has been corrected recently \cite{Margot2025}. The new parameters are much more reasonable because the coupling $t_4^{\perp}$ has been larger than $t_3^{\perp}$ leading to unrealistic transport properties. We developed additionally an effective low-energy Hamiltonian, which is based on \cite{Sousa2017} but with an empirically improved parameter $\lambda_n$. This model agrees well with the numerically calculated energy bands but we would like to emphasize that neither the original model nor our one are exact solutions of the Schrödinger equation. In \fig{4} we have demonstrated that the current flow depends strongly on the orientation of the underlying phosphorene lattice; while in one direction the current flow can be rather focused, it can be in the orthogonal direction disperse (using otherwise identical parameters). The low-energy current is concentrated mainly in the central layers, which can be understood by the vertical confinement of the electrons.

In a FLBP pn junction, generated by the electrostatic potential of a gate in a certain region of the system, see \fig{5}, the strong anisotropy of the current flow is even more prominent. The current passes the junction if its interface is aligned parallel to the zigzag direction of the phosphorene lattice, but it is reflected almost completely if the junction is aligned parallel to the armchair edge, see \fig{6}. This omni-directional total reflection, named anti-super-Klein tunneling, is due to opposite pseudo-spins in the different regions. In \fig{8}, we demonstrated that this effect appears for a rather wide range of electrostatic potential, provided that the electrons go from the valence to the conduction band (or vice versa). Finally, we proved in \fig{9} the robustness of the observed transport phenomena against oxidation of the top layer. As in the synthetization of FLBP is much more feasible than phosphorene, we believe that our research will stimulate further experiments and may pave the way towards electronic devices using FLBP.

\section{Acknowledgments}
JALB gratefully acknowledges his postdoctoral fellowship from DGAPA-UNAM.
We gratefully acknowledge funding from UNAM-PAPIIT IG100725 and IA102125.

\section{Appendix I: Transmission in a phosphorene pn junction}

In \fig{10} we show the transmission in a phosphorene pn junction as a function of the electrostatic potential. The figure proves, in the same way as \fig{8} for FLBP, that anti-super-Klein tunneling is observed for a current flow in the $y$ direction (armchair direction), if the electrons go from the conduction to the valence band (or vice versa) at the interface of the pn junction.

\section{Appendix II: Calculating the energy bands of FLBP}

\begin{figure}[t]
  \centering
  \includegraphics[scale=0.6]{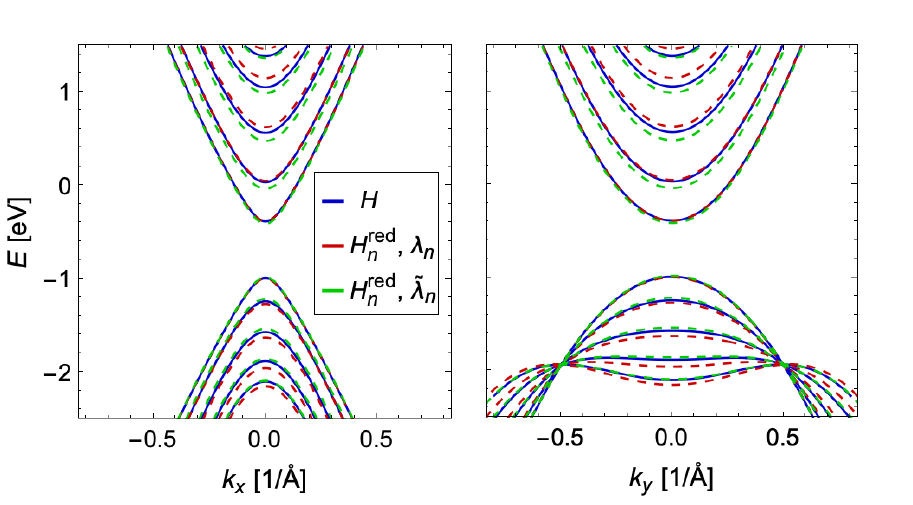}
  \caption{Energy bands of 5-layer FLBP calculated by means of the tight-binding Hamiltonian \eq{Hk} (blue curves), the reduced Hamiltonian \eq{Hr} with $\lambda_n$ (red dashed curve) and $\tilde\lambda_n$ (green dashed curve). The reduced Hamiltonian approximates well the tight-binding Hamiltonian.}
  \label{fig:11}
\end{figure}

After performing a plane-wave ansatz, the few-layer black phosphorous Hamiltonian in \eq{1} can be written in the form \cite{Sousa2017}
\begin{equation}
    H(\vec{k}) = \left(\begin{array}{c c c c c}
        H_\textrm{1L}(\vec{k}) & C(\vec{k}) & 0 & 0 & 0  \\
         C^\dagger (\vec{k}) & H_\textrm{1L}(\vec{k}) & C(\vec{k}) & \cdot & \cdot\\
         \cdot & C^\dagger(\vec{k}) & \cdot & \cdot & \cdot \\
         \cdot & \cdot & \cdot & \cdot & C(\vec{k}) \\
         0 & 0 & 0 & C^\dagger(\vec{k}) & H_\textrm{1L}(\vec{k})
    \end{array}\right)
    \label{eq:Hk}
\end{equation}
where the monolayer Hamiltonian is given by
\begin{equation}
    H_\textrm{1L}(\vec{k}) = 
    \left(\begin{array}{cc}
          H_1(\vec{k}) & H_2(\vec{k}) \\
          H_2(\vec{k}) & H_1(\vec{k})
    \end{array}\right), 
\end{equation}
with 
\begin{equation}
     H_1(\vec{k}) = 
     \left(\begin{array}{cc}
     t_{AA}(\vec{k})   & t_{AB}(\vec{k}) \\
     t^{*}_{AB}(\vec{k})  & t_{AA}(\vec{k})
    \end{array}\right),
\end{equation}
and
\begin{equation}
     H_2(\vec{k}) = 
     \left(\begin{array}{cc}
      t_{AD}(\vec{k}) & t_{AC}(\vec{k}) \\
      t^{*}_{AC}(\vec{k}) & t_{AD}(\vec{k})
    \end{array}\right).
\end{equation}
The matrix 
\begin{equation}
     C(\vec{k}) = 
     \left(\begin{array}{cc}
     0 & H_3(\vec{k}) \\
     0 & 0
    \end{array}\right),
\end{equation}
with
\begin{equation}
    H_3(\vec{k}) = 
    \left(\begin{array}{cc}
      t_{AD}^{\perp}(\vec{k})   & t^{\perp}_{AC}(\vec{k}) \\
      t^{\perp *}_{AC}(\vec{k})  & t^{\perp}_{AD}(\vec{k})
    \end{array}\right).
\end{equation}
represents the coupling between the layers. The coupling elements read
\begin{align}
    t_{AA}(\v{k})&= 2t_3 \cos(\v{k} \cdot \v{r}_3),\nonumber\\
    t_{AB}(\v{k})&= t_1 \textstyle \sum_{\v{r}_1} e^{\I \v{k} \cdot \v{r}_1},\nonumber\\
    t_{AC}(\v{k})&= t_2 e^{\I \v{k} \cdot \v{r}_2} +t_4 e^{\I \v{k} \cdot \v{r}_4},\nonumber\\
    t_{AD}(\v{k})&= 0,\nonumber\\
    t_{AC}^\perp(\v{k})&= t_1^\perp \textstyle \sum_{\v{r}_1^\perp}
    e^{\I \v{k} \cdot \v{r}_1^\perp} +t_4^\perp \textstyle \sum_{\v{r}_4^\perp} e^{\I \v{k} \cdot \v{r}_4^\perp},\nonumber\\    
    t_{AD}^\perp(\v{k})&= 2t_2^{\perp} \cos(\v{k} \cdot \v{r}_2^{\perp}) +t_3^{\perp} \textstyle \sum_{\v{r}_3^\perp}
    e^{\I \v{k} \cdot \v{r}_{3}^\perp}.    
\end{align}
The coupling constants con be found in Table I and the vectors to the $n$th nearest neighbors read
\begin{align}
\v{r}_1&= a_1(-\cos(\alpha/2),\pm \sin(\alpha/2)), \nonumber\\
\v{r}_2&= a_2(\cos(\beta),0),\nonumber\\
\v{r}_3&= a_1(0,2\sin(\alpha/2)),\nonumber\\
\v{r}_4&= (-(2a_1 \cos(\alpha/2)+a_2\cos(\beta),0),\nonumber\\
\v{r}_1^\perp&= (a_2 \cos(\beta), \pm a_1 \sin(\alpha/2)),\nonumber\\
\v{r}_2^\perp&= (a_1 \cos(\alpha/2) +a_2\cos(\beta), 0),\nonumber\\
\v{r}_3^\perp&= (\pm (a_1 \cos(\alpha/2) +a_2\cos(\beta)), \pm 2a_1 \sin(\alpha/2)),\nonumber\\
\v{r}_4^\perp&= (-(2a_1\cos(\alpha/2)+a_2\cos(\beta)), \pm a_1\sin(\alpha/2)).
\end{align}
The eigenvalues of the Hamiltonian in \eq{Hk} for a 5-layer FLBP system are shown in \fig{2} and \fig{11} (blue curves). By following the development in Ref. \cite{Sousa2017}, which consists of applying Toeplitz theory, the Hamiltonian $H(\v{k})$ can be decomposed \textit{approximately} to block-diagonal matrix form. Each $2 \times 2$ block matrix is given by
\begin{equation}
    \label{eq:Hr}
    H_{n}^{\text{red}}(\vec{k}) = H_1(\vec{k}) + H_2(\vec{k})+\lambda_n H_3(\vec{k}).
\end{equation}
We found empirically that the coefficient $\lambda_n \equiv \cos\bigl(\frac{2n-1}{2M+1}\pi\bigr)$ shows (slightly) better agreement as the expression found in Ref. \cite{Sousa2017}, $\tilde\lambda_n \equiv \cos\bigl(\frac{n}{M+1}\pi\bigr)$, compare the red and green dashed curves in \fig{11}. However, \eq{Hr} is an approximation and both values of $\lambda_n$ deviate from the energy bands of the tight-binding Hamiltonian.

It is known that any two-level system can be depicted by a general Hamiltonian
\begin{equation}\label{H2l}
    H_{2l} = d_0\sigma_0 + \vec{\sigma}\cdot\vec{d},
\end{equation}
where $\sigma_0$ and $\vec{\sigma} = (\sigma_x,\sigma_y,\sigma_z)$ are the Pauli matrices, $d_0$ and $\vec{d} = (d_x,d_y,d_z)$ indicates the pseudo-spin direction in the Bloch sphere. Eigenvalues and eigenvectors are
\begin{equation}
    E_s = d_0 + s d,
\end{equation}
with the band index $s = \pm 1$, and
\begin{eqnarray}
    \vec{\Psi}_{s = 1} & = & (\cos\theta/2, \textrm{e}^{i\phi}\sin\theta/2), \nonumber\\
    \vec{\Psi}_{s = -1} & = & (\sin\theta/2, -\textrm{e}^{i\phi}\cos\theta/2),
\end{eqnarray}
where $(\theta,\phi)$ are the spherical coordinates of $\vec{d}$ in the Bloch sphere. Comparing Eqs. \eq{Hr} and \eqref{H2l}, we obtain
\begin{eqnarray}
    d_0(\vec{k}) & = & t_{AA}(\vec{k})+t_{AD}(\vec{k})+\lambda_n t_{AD}^\perp(\vec{k}), \nonumber\\
    d_x(\vec{k}) & = & \Re{t_{AB}(\vec{k})+t_{AC}(\vec{k})+\lambda_n t_{AC}^\perp(\vec{k})}, \nonumber\\
    d_y(\vec{k}) & = & -\Im{t_{AB}(\vec{k})+t_{AC}(\vec{k})+\lambda_n t_{AC}^\perp(\vec{k})}, \nonumber\\
    d_z(\vec{k}) & = & 0.
\end{eqnarray}
Since $d_z(\vec{k}) = 0$ then $\theta = \pi/2$, the pseudo-spin is oriented in the $xy$ plane with the angle $\phi = \arctan(d_y/d_x)$.

For a continuum model, we expand the Hamiltonian around the $\Gamma$ point up to second order in $\vec{k}$, ending up with the Hamiltonian $H_n^\text{red}$ in \eq{4}.

\bibliography{flbp}

\end{document}